# Laser-beam patterned topological insulating states on thin semiconducting MoS$_2$

H. Mine[1], A. Kobayashi[1], T. Nakamura[2], T. Inoue[3], S. Pakdel[5,6,9], E. Z. Marin[7], D. Marian[7], E. Gonzalez-Marin[7], S. Maruyama[3], S. Katsumoto[2], A. Fortunelli[8], J. J. Palacios[4,5], J. Haruyama[1,2*]

[1]*Faculty of Science and Engineering, Aoyama Gakuin University, 5-10-1 Fuchinobe, Sagamihara, Kanagawa 252-5258, Japan.*
[2]*Institute for Solid State Physics, The University of Tokyo, 5-1-5 Kashiwanoha, Kashiwa, Chiba 277-8581, Japan.*
[3]*Dept Mechanical Engineering, The University of Tokyo, 7-3-1 Hongo, Bunkyo-ku, Tokyo 113-8656, Japan.*
[4]*Department of Physics, The University of Texas at Austin, Austin, Texas 78712, USA.*
[5]*Departamento de Física de la Materia Condensada, Instituto Nicolás Cabrera (INC), and Condensed Matter Physics Center (IFIMAC), Universidad Autónoma de Madrid, E-28049 Madrid, Spain.*
[6]*School of Electrical and Computer Engineering, University College of Engineering, University of Tehran, Tehran 14395-515, Iran.*
[7]*Dipartimento di Ingegneria dell'Informazione, Università di Pisa, Pisa 56122, Italy*
[8]*CNR, National Research Council, 56124 Pisa, Italy*
[9]*Department of Physics and Astronomy, Aarhus University, 8000 Aarhus C, Denmark*

[*]To whom correspondence should be addressed. E-mail: J-haru@ee.aoyama.ac.jp

**Identifying the two-dimensional (2D) topological insulating (TI) state in new materials and its control are crucial aspects towards the development of voltage-controlled spintronic devices with low power dissipation. Members of the 2D transition metal dichalcogenides (TMDCs) have been recently predicted and experimentally reported as a new class of 2D TI materials, but in most cases edge conduction seems fragile and limited to the monolayer phase fabricated on specified substrates. Here, we realize the controlled patterning of the 1T'-phase embedded into the 2H-phase of thin semiconducting molybdenum-disulfide (MoS$_2$) by laser beam irradiation. Integer fractions of the quantum of resistance, the dependence on laser-irradiation conditions, magnetic field, and temperature, as well as the bulk gap observation by scanning tunneling spectroscopy and theoretical calculations indicate the presence of the quantum spin Hall phase in our patterned 1T' phases.**

Two-dimensional (2D) topological insulting (TI) states have been mainly investigated in HgTe/CdTe or InAs/GaSb quantum well systems [1-3]. In the 2D TI state the quantum spin Hall (QSH) effect emerges thanks to the simultaneous presence of a bulk energy gap and gapless helical edge states protected by time-reversal symmetry, namely, opposite and counter-propagating spin states forming a Kramers doublet. Interestingly, 2D TI states were first theoretically predicted for graphene [4-6], but experimentally reported in only few related systems [7-9] such as low-coverage Bi$_2$Te$_3$ nanoparticle-decorated graphene [8]. Moreover, control of the QSH phase in graphene-based systems remains a challenge.

Recently, a family of atom-thin transition metal dichalcogenides (TMDCs) materials has also been predicted to exhibit the QSHE [10-12], having its origin in the natural band inversion of the 1T' phase (one of the phases of TMDC; see Supplementary Material (SM) 1) and the spin-orbit coupling (SOC)-induced band-gap opening. Moreover, the TI state has been experimentally verified in the case of WTe$_2$ [13-15] thanks to the stability and high-quality of WTe$_2$ monolayers carefully formed on bi-layer graphene/atom-thin hBN. Various signatures of the TI state have been demonstrated in this material [13, 15], including the latest observation of a half-integer quantum value of resistance ($R_Q/2 = h/2e^2 = 12.9$ kΩ, where $h$ is Planck's constant and $e$ is the charge on the electron) [14].

However, the TI phenomenon in WTe$_2$ is rather sensitive to the substrates, synthesis process, and the chemical environment, making its controlled use in practical applications challenging. Moreover, although the (metastable) 1T' phase can be found or induced in other TMDCs [16,17], nobody has demonstrated the existence of the QSHE in these other TMDCs except for [18]. The conditions under which helical edge states can exist at the 1T'-2H interfaces is a crucial problem which should be mastered for both TI physics and its applications. Here, we pattern a metallic 1T'-phase (SM 1) embedded into the non-topological and semiconducting 2H phase of thin MoS$_2$ flakes, one of the TMDC family, via a facile protocol based on laser beam irradiation [16] and, via transport measurements, reveal that the embedded 1T' phase exhibits the TI state.

In the present experiments thin MoS$_2$ flakes are obtained by mechanical exfoliation of the bulk material and transferred onto a SiO$_2$/Si substrate. Layer thicknesses ~17 nm have been confirmed by atomic force microscopy (AFM) and optical microscopy (OM). OM and AFM images of a flake with two different patterns created by laser beam irradiation are shown in Figs. 1A and 1B (SM 2). The large rectangular pattern (with two electrode probes) is analogous to that previously used for graphene under high-magnetic fields [9] and for monolayer WTe$_2$ [14]. In contrast, the H-letter like pattern is analogous to that used in HgTe/CdTe quantum wells [1] and in our previous Bi$_2$Te$_3$-nanoparticle decorated graphene system [8]. The presence of helical edge states was confirmed in all these systems. In the OM image, the color of the irradiated patterns drastically changes to semi-transparent (Fig. 1(a)). The cross-sectional AFM image of the irradiated part reveals a decrease in the thickness of about 10 nm (Fig. 1(b)). These observations are consistent with previous reports in multilayer MoTe$_2$ (16), which demonstrated a layer thinning effect caused by the burn-out of individual monolayers due to in-plane heat accumulation from the laser beam irradiation. Heat introduced by the laser irradiation causes, in turn, a 2H-1T' transition in the upper layers (Fig. 1(c)) ([16]; SM 3).

Typical Raman spectra are shown in Fig. 1(d) and we can use it to estimate the number of 1T' layers formed. For the non-laser-irradiated region (2H phase), the Raman peaks are evident for multi-layer (> five layers) MoS$_2$, showing the large and characteristic E$_{2g}$ and A$_{1g}$ peaks, while the pattern of the peaks for the laser-irradiated region (1T'+2H) has definitely changed. Both peaks are still visible, which can be attributed to the 2H layers remaining underneath the 1T' layers (SM 3), while the other peaks can be certainly attributed to the 1T' phase. Because the thickness of the laser-irradiated part is ~ 7 nm, as mentioned above, and the minimum thickness expected for the remaining 2H layers underneath the 1T' ones is 5 layers × ~0.7 nm = ~3.5 nm (the minimum thickness to get Raman signal), we estimate a maximum thickness for the 1T' layers of ~3.5 nm. Since we expect the laser irradiation to slightly damage the top-most 1T' layers, the number of pristine 1T' layers is expected to be at most a few.

Photo luminescence (PL) signals of the laser-irradiated parts are shown in Figs. 1(e) and 1(f). These reveal that the peak positions shifted to lower wavelengths (*i.e.*, higher energies) and the peak intensities decreased with laser-irradiation time. This is also compatible with the above-mentioned interpretation. When the upper layers are transformed into the 1T' phase, these layers cause no PL signals [19], whilst the remaining bottom semiconducting layers still yield a (reduced) signal (Fig. 1(c)). Furthermore, X-ray photoelectron spectroscopy (XPS) of the laser-irradiated part demonstrates the two types of hybridization of Mo 3d orbitals associated with the 1T' and 2H phases (Fig. 1(g)).

Resistance ($R$) measurements as a function of back gate voltage ($V_{bg}$) of the samples shown in Fig. 1(a) are shown in Fig. 2. Au/Ti electrodes are in contact only with four corners of the rectangular pattern and each branch of the *H*-letter-like patter so as to measure charge-spin transport only in the 1T' region (insets of 2(a) and 2(c)). Because of the presence of a Schottky barrier at the 1T'/2H junction (> ~0.2 eV; see Fig. 3(c) for the semiconducting band gap in 2H region) [19,20], the 2H semiconducting layers below the 1T' layers are expected to give a negligible contribution to these measurements, particularly at low temperatures.

For the rectangular pattern, the two-terminal resistance between electrodes 1,3 and 2,4 is measured as a function of $V_{bg}$ by flowing a constant current between electrodes 1,3 and 2,4 (see insets). The results of two samples formed after different irradiation times to each point (20 s and 30 s) are shown in Figs. 2(a) and 2(b), respectively. Individual figures demonstrate two *R* peaks (SM 10). At high $V_{bg}$, *R* peaks of $R_{34} \approx R_Q/2$ are confirmed in both figures (at $V_{bg} \approx$ +25V and +19V in Figs. 2(a) and 2(b), respectively). Larger *R* peaks are also observed at negative $V_{bg}$'s ($R_{34} \approx R_Q$ and $\approx 3R_Q/2$ at $V_{bg} \approx$ -5V and ≈ -10V in Figs. 2(a) and 2(b), respectively). For the *H*-letter like pattern, when a constant current flows between electrode pair 1-2, the non-local resistance ($R_{NL}$) between electrode pair 3-4 ($R_{34}$) is measured as a function of $V_{bg}$ (see inset of Fig. 2(c)). Figure 2(c) shows the result. A $R_{NL}$ plateau of $R_{34} \approx R_Q/4$ at high $V_{bg}$ (≈ +20-25V) is confirmed with a $R_{NL}$ peak of $R_{34} > R_Q/2$ at low $V_{bg}$ (≈ +5V) also appearing. As usually occurs, large $V_{bg}$ values need to be applied to significantly tune the Fermi level via the SiO$_2$/Si substrate (SM 4) [21].

The $R_Q/2$ and $R_Q/4$ of *R* peak values observed at high $V_{bg}$ suggests the appearance of the QSH phase without and with dephasing in metal electrodes, respectively [1,8]. First, the *R* peak values ~ $R_Q/2$ confirmed for the 1T' rectangular patterns (Figs. 2(a) and 2(b)) are consistent with the presence of helical edge modes without dephasing. In this case, the two counter-propagating spin channels can be preserved at two different quasi-chemical potentials between the electrodes, leading to a net current flow along the edges with *R* equal to $R_Q/2$ (as based on the Landauer-Büttiker (LB) formalism). Such a two-terminal resistance plateau ~ $R_Q/2$, reported in both rectangular mono-layer WTe$_2$ [14] and in large rectangular graphene under high magnetic fields with a similar electrode connection to the present one (insets of Fig. 2(a,b)) [9] was presented as evidence for helical edge states [22]. Ref. (*14*) also reported deviations from ~ $R_Q/2$ to larger values in long-channel samples (>> a few 100nm). Remarkably, in our case, *R* does not deviate significantly from $R_Q/2$ even for ~1 μm channel lengths, because of the possibly high uniformity of the 1T' phase formed by our highly uniform laser beam (SM 5).

Second, the $R_{NL}$ plateau value ~ $R_Q/4$ (Fig. 2(c)) corresponds to the case of helical edge modes with dephasing in the metal electrodes in the *H*-letter like pattern with four metal electrodes (inset) and nicely agrees with the result in Refs. [1,8]. Once the helical edge electrons enter the voltage electrodes, they interact with a reservoir containing an infinite number of low-energy degrees of freedom so that time-reversal symmetry is effectively broken by the macroscopic irreversibility. In particular, a resistance plateau can be confirmed in Fig. 2 (c). The plateau shape, as opposed of a peak, is attributed to the channel width being narrower (~1 μm) than that in the rectangular pattern (~2 μm) and the usage of highly uniform laser beam as mentioned above (SM 6). This is also consistent with the results in [1]. Consequently, the two counter propagating channels equilibrate at the same chemical potential, determined by the voltage of the metallic electrodes, leading to dissipation and, thus, emergence of integer fractions of $R_Q$. Therefore, these *R* peaks (*i.e.*, ~ $R_Q/2$ in the rectangular pattern and ~$R_Q/4$ in the *H*-letter like pattern) evidently suggest the presence of the helical edge states in the patterned 1T' layers. These *R* peaks without influence of dephasing will originate from either monolayer or a few layers existing below the top-surface layer, while it is speculated that the other *R* peaks (~ $R_Q$ and ~$3R_Q/2$ in Figs. 2(a) and 2(b), respectively), which are typically larger than those observed at high $V_{bg}$, are derived from the top-surface 1T' layer with edge defects (SM 7,8) [24].

Indeed, we observe single $R$ peak with $R_Q/2$ value, which is consistent with a conventional QSH phase [14], when lower-power laser is irradiated (~4.6 mW) (Fig. 2(d)). This supports the abovementioned argument that the observed two $R$ peaks with the large $R$ peak values (Figs. 2(a-c)) are attributed to the excess heat accumulation through the multi-layers caused by the present high-power laser irradiation. The low-power laser irradiation suppresses the excess heat accumulation and allows a topological transition in mono-layer with less edge defects (SM3).

Perpendicular magnetic-field ($B_\perp$) dependence measurements for the samples shown in Figs. 2(b) and 2(c) are demonstrated in Fig. 3(a). As $B_\perp$ increases, the conductance $G$, corresponding to the inverse of three $R$ peak values in Fig. 2(b,c), exponentially decrease. $G$ values corresponding to the inverse of two off-$R$ peaks (Fig. 2(b)) remain almost unchanged. These results are in good agreement with those in the QSHE observed in $WTe_2$ [14] and our $Bi_2Te_3$ decorated graphene [8], and evidently support that the first three $R$ peaks can be attributed to helical edge states. Only when the Fermi level is set to the Kramers degeneracy point, $R$ values can well reflect the band gap opening due to Zeeman effect caused by applied $B_\perp$ (see insets), resulting in the observed exponential decrease in $G$. The pink symbol observed in the $H$-letter like pattern show a saturation at lower $B_\perp$ (~3 T), because the channel width (~1 μm) is smaller than the rectangular pattern and other edge effects may disturb the exponential $G$ decrease.

As far as zero-$B_\perp$ temperature dependence is concerned, $G$ corresponding to inverse of the three $R$ peak values ($R_Q/2$, $3R_Q/2$, and $R_Q/4$) remain constant up to the transition temperatures ($T_{c1}$) of ~40 K (red symbol), ~25 K (blue), and ~30 K (pink), respectively (Fig. 3(b)). At temperatures higher than the individual $T_{c1}$, $G$ increases as temperature increases, following the thermal activation formula (*i.e.*, a linear dependence in the Arrhenius plot of Fig. 3(b)) with activation energies of ~15 meV (red symbol), ~10 meV (blue), and ~7 meV (pink), respectively. Moreover, $G$ increases again above $T_{c2}$ ~60 K (red), ~50 K (blue), and ~60 K (pink), respectively, with larger slopes. The $T_{c1}$, which are lower than 100 K as reported in $WTe_2$ [14], are attributed to bulk gaps being smaller than that of $WTe_2$ [25]. In contrast, $G$ increases at temperatures > $T_{c2}$ have not been observed previously [8, 14]. This can be attributed to the thermally activated carriers flowing into the 1T' QSH phase region from the 2H semiconducting region over the Schottky barrier (or band discontinuity) at the 1T'/2H layer interface, because the barrier height is much larger than the bulk gaps [20]. The largest activation energy (i.e., barrier height or band discontinuity) for the pink symbol is consistent with the smallest gap value of ~7 meV as mentioned above.

Scanning tunneling spectroscopy (STS) spectra of the non-laser-irradiated 2H region and the irradiated rectangular 1T'-pattern are shown in Fig. 3(c-e) (SM 9). For (c), it demonstrates evident gaps ~0.6 eV, which are in good agreement with n-type semiconducting gap of thin $MoS_2$ with number of the layers larger than five. This gap is large enough to embed the present topological gaps with ~10 meV order. For (d) with $V_{bg}$ tuned to Kramers point, in the two bulk points, STS gaps of ~25 - 35 meV are confirmed, while the gap disappears at an edge. The bulk gap values almost agree with the values estimated from the temperature dependence of resistance peaks as mentioned above. Although they are smaller than the 45 meV gap reported in 1T'-$WTe_2$ [15], they are appropriate for 1T'-$MoS_2$. For (e), as $V_{bg}$ runs away from Kramers point (*i.e.*, $V_{bg}$ ~ -10V) (see insets), the d$I$/d$V$ increases and the d$I$/d$V$ dip disappears, resulting in just a metallic behavior of 1T' phase. This supports that disappearance of the bulk gap in (d) is attributed to edge current due to Kramers point. Consequently, all results suggest that the laser-created 1T' phases can be in QSH phases with helical edge modes.

The existence of helical edge states at the 1T'/2H interface is further supported by our theoretical calculations [26]. Fig. 4(a) shows the overall band structure of the heterostructure. Our choice of a centrosymmetric system with two interfaces makes all the bands doubly-degenerate. One can clearly identify the bulk band inversion associated with the 1T' phase near $\Gamma$ and the gap opened by the SOC (enhanced by the quantum confinement due to the finite width of the 1T' region). Importantly, we learn from these calculations that the 1T' phase gap is in the middle of the much larger gap of the 2H phase (indicated by a yellow shade), which should enable the manifestation of the protected helical states along the interfaces. Fig. 4(b) shows a zoom of the band structure in the relevant low-energy sector in half of the Brillouin zone. Despite the complexity of the band structure, which stems from the presence of native trivial edge states of the zigzag edges of the 2H phase [27], the total number of crossings at any energy within the gap in half of the Brillouin zone is odd, as expected.

Fig. 4(c) shows the band structure of a bilayer for different stacking geometries (shown in the insets). These three geometries do not exhaust all stacking possibilities, but are representative of those found in bulk materials. After relaxation, a gap remains open in all cases [28], which is a necessary condition for the observation of the QSH phase. Also, the band structure reflects the weak coupling between layers, which also facilitates the observation of the QSH phase.

Controlled patterning of the 1T'-phase onto the 2H-phase of thin semiconducting $MoS_2$ by laser beam irradiation has been demonstrated. Using multi-layers brings many advantages to the present laser-beam irradiated experiments (*e.g.*, protection from laser damage and oxidation) (SM 11). Further optimization of the conditions for laser irradiation will allow for on-demand patterning of 2D (or 1D) topological phases onto desired positions of non-topological phases of TMDCs and will facilitate the path towards topological quantum computation [29].

**ACKNOWLEDGEMENTS**

The authors thank S. Tang, Z.-X. Shen, A. MacDonald, S. Murakami, Y. Shimazaki, T. Yamamoto, S. Tarucha, T. Ando, R. Wu, J. Alicea, M. Dresselhaus, P. J.-Herrero, and P. Kim for their technical contributions, fruitful discussions, and encouragement. The work carried out at Aoyama Gakuin University was partly supported by a grant for private universities and a Grant-in-Aid for Scientific Research (15K13277) awarded by MEXT. The work at the University of Tokyo was partly supported by Grant-in-Aid for Scientific Research (17K05492, 18H04218, 25247051, and 26103003). JJP and SP acknowledge Spanish MINECO through Grant FIS2016-80434-P, the Fundación Ramón Areces, the María de Maeztu Program for Units of Excellence in R&D (MDM-2014-0377), the Comunidad Autónoma de Madrid through NANOMAGCOST Program, and the European Union Seventh Framework Programme under Grant agreement No. 604391 Graphene Flagship. SP acknowledges the computer resources and assistance provided by the Centro de Computación Científica of the Universidad Autónoma de Madrid. DM and EG-M gratefully acknowledge the support from the Graphene Flagship Graphene Core2 Contract No. 785219. DM, EG-M, and AF are very grateful to Gianluca Fiori and Giuseppe Iannaccone for continuous support, strong encouragement, and enlightening discussions.

relaxed atomic positions of the unit cells near the interfaces (detail shown in Fig. 4(b)) are obtained from density functional theory (DFT) structural relaxations, in which we have kept a flat structure away from the interfaces as to simulate the underlying $MoS_2$ layers. We have corroborated our modeling by investigating other atomistic models of the 2H/1T' interface, starting from different initial configurations, also considering an asymmetric periodic model, and relaxing them using different constraints (see SM12,13, which include Refs. [30-40]). The DFT results in all these investigated cases converged to a qualitatively similar picture in terms of structural stability and band structure, which supports the robustness of the model here proposed.

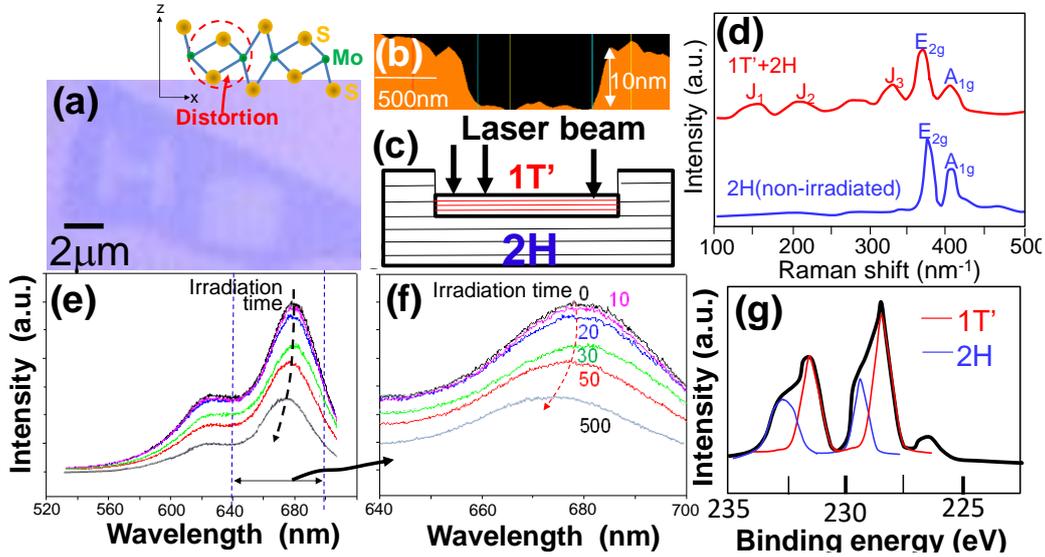

**Fig. 1 (a)** Optical microscopy image of the 1T'-phase rectangular (right) and *H*-letter-like (left) patterns formed onto a thin 2H-MoS$_2$ flake by laser beam irradiation (see methods). **Inset**: Schematic cross section of a crystal structure of 1T'-MoS$_2$ mono-layer with distortion. **(b)** AFM image of a cross section of the laser irradiated part. **(c)** Schematic cross section of 1T' phase part created by laser-beam irradiation onto few-layer MoS$_2$, corresponding to (b). **(d)** Example of Raman spectra for non-laser irradiated region (2H phase; blue curve) and irradiated region (1T' phase on 2H phase; red curve). Individual peaks correspond to E$_{2g}$ ~382 nm$^{-1}$ and A$_{1g}$ ~408 nm$^{-1}$ for 2H phase, J$_1$ ~155 nm$^{-1}$, J$_2$ ~225 nm$^{-1}$, and J$_3$ ~330 nm$^{-1}$ for 1T' phase. **(e,f)** PL spectra of the laser-beam irradiated points plotted for wavelengths **(e)** 530-710 nm and **(f)** 640–700 nm. The numbers on the graph are the irradiation time for each plotted line and are common to both (e) and (f). **(g)** X-ray photoelectron spectroscopy (XPS) of the sample after laser irradiation. Red and blue lines are data fits for spectra of the 1T' and 2H phases, respectively.

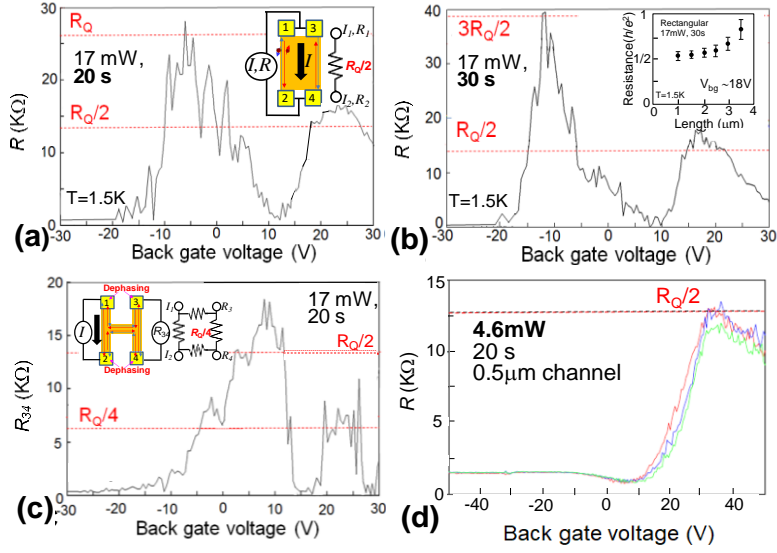

**Fig. 2 (a,b)** For the 1T'-rectangular patterns formed by two different laser irradiation times on each points ; two-terminal resistance measured between electrodes 1,3 and 2,4 as a function of $V_{bg}$ by flowing a constant current between electrodes 1,3 and 2,4 (**insets**). Contact resistances with 1T' metallic-layer resistances are subtracted. **(c)** For the 1T' *H*-letter like pattern; Non-local resistance ($R_{NL}$) observed for electrode pairs 3-4 as a function of $V_{bg}$, when a constant current flow between electrode pairs 1-2 (**inset**). **(d)** For the 1T'-rectangular pattern formed by reduced laser power, with a short channel. Three different colors correspond to three-time measurements. Equivalent circuits are shown in insets of (a,b). **Inset of (b)**; Channel length dependence of *R* peak values in high $V_{bg}$ regions. Error bars are for the results of each three samples.

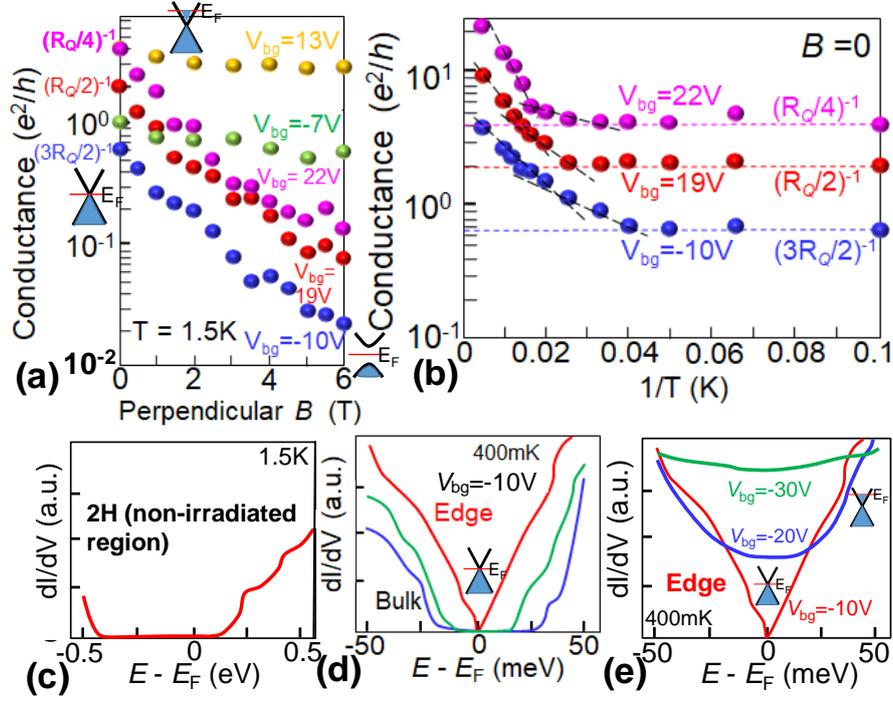

**Fig. 3 (a)** Out-of-plane magnetic-field ($B_\perp$) dependence of conductance corresponding to inverse of the three $R$ peaks (blue and red symbols for Fig. 2(b) (rectangular pattern) and pink for high $V_{bg}$ of Fig. 2(c) (H-letter like pattern)) and two off-$R$ peak values (green and orange symbols for Fig. 2(b)). **(b)** Temperature dependence of conductance corresponding to the three $R$ peaks in (a) in Alenius plot format. Dashes lines are guide to eyes. **(c-e)** STS spectra for non-laser-irradiated 2H region **(c)** and irradiated 1T' region **(d,e)** (SM 9); **(d)** The two bulk signals (blue and green lines) were measured near the center of the 1T'-rectangular pattern, and the edge signal (red line) was measured near the boundary of the 1T'/2H phases. **(e)** The edge signals for different three $V_{bg}$ values in sample of (d). The 1T' region was formed by irradiation with the condition for Fig. 2(b). Insets through all figures are schematic views of band diagram near Kramers point with Fermi level ($E_F$).

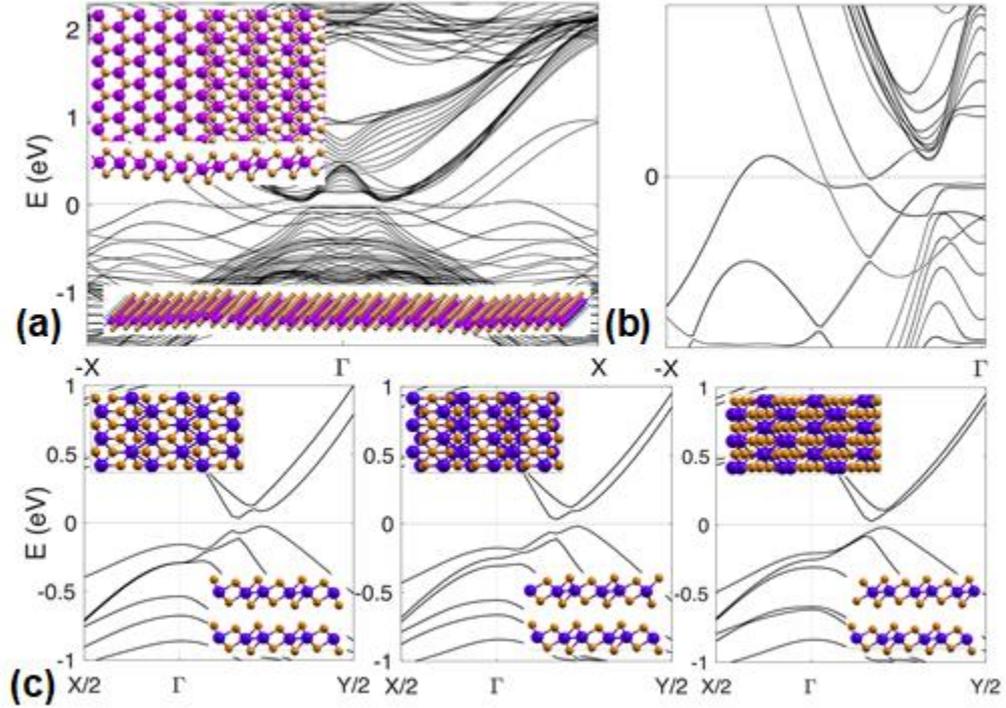

**Fig. 4**. **(a)** Overall band structure of the 2H/1T'/2H heterostructure (bottom inset: a view of the whole heterostructure showing the passivated edges; top inset: atomic detail of one interface). The yellowish area indicates the energy window within the 2H phase gap. Inside this range and near the center of the Brillouin zone, the bulk band inversion of the 1T' phase and the gap opened by the SOC can be seen along with non-trivial and trivial bands. **(b)** A zoom of the bands into the relevant energy window. The number of band crossings at the Fermi level (placed at zero) and at any energy in the gap is odd, as expected from the presence of protected interface states. **(c)** Bulk band structure of a bilayer for three different stacking possibilities (indicated in the insets), showing a gap in all of them.